\documentclass[twocolumn,showpacs,preprintnumbers,amsmath,amssymb,pra,superscriptaddress]{revtex4}
\usepackage{graphicx}
\usepackage{epstopdf}
\usepackage{dcolumn}
\usepackage{bm}
\usepackage{amsmath}
\usepackage{amssymb}
\usepackage{mathrsfs}
\usepackage{amsfonts}
\usepackage{color}

\begin{document}

\title{Bose-Hubbard Model: Relation Between Driven-Dissipative Steady-States\\ and Equilibrium Quantum Phases}
\author{Alexandre Le Boit\'e}
\author{Giuliano Orso}
\author{Cristiano Ciuti}
\affiliation{Laboratoire Mat\'eriaux et Ph\'enom\`enes Quantiques,
Universit\'e Paris Diderot-Paris 7 and CNRS, \\ B\^atiment Condorcet, 10 rue
Alice Domon et L\'eonie Duquet, 75205 Paris Cedex 13, France}

\begin{abstract}

We present analytical solutions for the mean-field master equation of the driven-dissipative Bose-Hubbard model for cavity photons, in the limit of both weak pumping and weak dissipation.  
Instead of pure Mott insulator states, we find statistical mixtures with the same second-order coherence $g^{(2)}(0)$ as a Fock state with $n$ photons, but a mean photon number of $n/2.$
These mixed states  occur when $n$ pump photons have the same energy as $n$ interacting photons inside the nonlinear cavity and
survive up to a critical tunneling coupling strength, above which a crossover to classical coherent state takes place.
We also explain the origin of both antibunching and superbunching predicted by P-representation mean-field theory at higher pumping and dissipation. 
In particular, we show that the strongly correlated region of the associated phase diagram cannot be described within  
the semiclassical Gross-Pitaevski approach.

\end{abstract}
\pacs{42.50.Ar,03.75.Lm,42.50.Pq,71.36.+c}

\maketitle

\section{introduction}

Since the seminal work by Fisher {\it et al.} in 1989 \cite{Fisher}, the Bose-Hubbard model and its celebrated Mott insulator to superfluid quantum phase transition has attracted a great deal of attention. In spite of being the simplest model of interacting bosons on a lattice, its relevance for the study of many-body phenomena in bosonic systems has become even more prominent since the experimental observation of the predicted phase transition with ultracold atoms in optical lattices \cite{Greiner}. At the time of the first theoretical investigations, the most relevant candidates for experimental verification were undoubtedly atoms or Cooper pairs.  But with the tremendous experimental progress in quantum optics and quantum electrodynamics of the past twenty years \cite{Houck,Deveaud},  the exciting field of many-body physics has ceased to belong exclusively to the realm of genuine particles: the exploration of Bose-Hubbard physics in optical systems is now within reach. 
It has indeed been demonstrated that in a nonlinear optical medium, light behave as a quantum fluid of interacting photons \cite{RMP} and some of the most remarkable features of quantum fluids,  such as superfluid propagation \cite{Amo09,Tanese2012} or generation of topological excitations \cite{Pigeon, Amo11, Nardin,Sanvitto} have been observed in experiments with solid-state microcavities. It has also become possible to design arrays of nonlinear cavity resonators, such that the  effective on-site  photon-photon interactions are large enough to enter the strongly correlated regime \cite{Imamoglu, Fink, Liew, Bamba}.

The question of whether key features of equilibrium physics are still present when the bosons have a finite life time has been of crucial importance ever since the first theoretical proposals for implementing the Bose-Hubbard model with photons or polaritons \cite{Hartmann06, Greentree, Angelakis}. Exploiting the analogies between the two models,  early works were focused on phenomena close to the equilibrium Mott insulator-Superfluid transition \cite{Tomadin, Wu}. More recent studies were devoted to the interesting  driven-dissipative regime, where the cavity resonators are excited by a coherent pump which competes with cavity losses \cite{Carusotto09, Ferretti, Hartmann10, Leib10, Hafezi2011,Nissen,Carusotto2012, Joshi, Jin,Angelakis2013}. In such conditions, the role of dissipation is crucial and similarities with equilibrium phenomena are more difficult to uncover. In particular, our recent study of the 2D mean-field phase diagram in the thermodynamical limit showed that the system is driven into steady-state phases whose general properties and collective excitations may be, at first sight, very different from equilibrium \cite{PRL2013}. 

In this context, it is important to clarify the relation between driven-dissipative and equilibrium models and gain more insight into the nature of the stationary states. 
To tackle this question, we focus in this article on the limit of weak dissipation and weak pumping.  Analytical expressions for the density matrix and the relevant observables enable us to identify the non-classical states of light which are the most closely related to a Mott insulator. Such states may be prepared when multiphotonic absorption processes become resonant. 
We show that beyond a critical value of the hopping rate between neighboring cavities, these non-classical states disappear and the system is driven into a classical coherent state. These results cast light on the photon statistics observed for stronger pumping and stronger dissipation. In particular, they give a clear picture of the two phases presented in the bistability diagram of ref. \cite{PRL2013}. We also show that the `weakly-interacting' sector of such diagram is well understood by means of a Gross-Pitaevski approximation.  As expected, this simplified approach fails when on-site repulsion is much larger than the coupling between sites.

The paper is organized as follows: The model is introduced in section \ref{sec:model}. Section \ref{sec:WPE} is devoted to the limit of weak pumping and weak dissipation. The density matrix along with the relevant observables are first computed for  single cavity (\ref{subsec:single_cav}), and then extended to coupled cavities (\ref{subsec:coupled_cav}). In section \ref{sec:GP} we explore the Gross-Pitaevski regime and we conclude in section \ref{sec:conclu}.
%
%
\section{The Model}
\label{sec:model}

We consider a driven-dissipative Bose-Hubbard model under homogeneous coherent pumping describing a bidimensional square lattice of cavity resonators. 
The system is described by the following Hamiltonian \cite{RMP}:
\begin{align}
H =  &-\frac{J}{z}\sum_{<i,j>}b^{\dagger}_i b_j+\sum_i^{N}\omega_c b^{\dagger}_i b_i+\frac{U}{2}b^{\dagger}_i b^{\dagger}_i b_ib_i \nonumber  \\   
&+ Fe^{-i\omega_pt}b^{\dagger}_{i}+ F^{*}e^{i\omega_pt}b_{i}, \label{Hamilto}
\end{align}
 where $b^{\dagger}_i$ creates a boson on site $i$, $J > 0$ is the tunneling strength,  and $z=4$ is the coordination number. $<i,j>$ indicates that tunneling is possible only between first-neighbors. $U > 0$ represents the effective on-site repulsion, $F$ is the amplitude of the incident laser field, $\omega_c$  is the bare cavity frequency and $\omega_p$ the frequency of the pump. The dynamics of the many-body density matrix $\rho(t)$ is described in terms of a Lindblad master equation:
\begin{equation}
i\partial_{t} \rho = [H, \rho] + \frac{i\gamma}{2}\sum_{i}^{N} 2b_{i} \rho b_{i}^{\dagger} - b_{i}^{\dagger} b_{i} \rho - \rho  b_{i}^{\dagger} b_{i}, \label{ME}
\end{equation}
where $\gamma$ is the dissipation rate. It is convenient to eliminate the time dependency in Eq.(\ref{Hamilto}) by performing a unitary transformation on the density matrix:
\begin{equation}
\rho \rightarrow U\rho U^{\dagger},
\end{equation}
where $U = e^{i\omega_pt\sum_i b^{\dagger}_ib_i}$. This amounts to writing the density matrix in a frame rotating at the pump frequency $\omega_p$. In this rotating frame, the Hamiltonian governing the dynamics is now time independent:
\begin{equation}
H_{rf} = -\frac{J}{z}\sum_{<i,j>}b^{\dagger}_i b_j-\sum_i^{N}\Delta \omega b^{\dagger}_i b_i+\frac{U}{2}b^{\dagger}_i b^{\dagger}_i b_ib_i + Fb^{\dagger}_{i}+ F^{*}b_{i}. \label{Hamilto_rf}
\end{equation}
We have introduced the detuning between the pump the bare cavity frequency $\Delta \omega = \omega_p - \omega_c$. While for equilibrium quantum gases, the chemical potential $\mu$ is a key quantity, in this non-equilibrium model the steady-state phases depend instead on the pump parameters
$F$ and $\Delta \omega$, which compete with  $\gamma$. It is worth pointing out that this is a model which well describes a lattice of cavities whose extra-cavity environment is the electromagnetic vacuum (apart from the applied driving field).

Apart from being conceptually simple and rich in phenomenology, the Bose-Hubbard model has another advantage: mean-field theory gives very good qualitative results at equilibrium. This approach has thus been extended to the driven-dissipative model under homogeneous pumping \cite{PRL2013}. It consists of a decoupling approximation $b^{\dagger}_{i}b_{j} \to \langle b^{\dagger}_i \rangle b_j+\langle b_j \rangle b^{\dagger}_i$, that reduces the initial $N$-site Hamiltonian to an effective single-site problem:
%
\begin{equation}
H_{mf} = -\Delta \omega b^{\dagger} b+\frac{U}{2}b^{\dagger} b^{\dagger} bb + (F-J\langle b \rangle)b^{\dagger}+ (F^{*}-J \langle b \rangle ^{*})b, \label{H_MF}
\end{equation}
the effective external coherent field being:
\begin{equation}
F \rightarrow F' = F-J\langle b \rangle \label{MF},
\end{equation}
where the bosonic coherence $\langle b \rangle$ has to be determined self-consistently. An exact solution of the single-cavity problem may be obtained by using the complex $P$-representation for the density matrix \cite{PRL2013, DW, Vidanovic}. In this phase space approach, the matrix $\rho$ is expressed in a coherent state basis as:
\begin{equation}
\rho = \int_{\mathcal{C}_{\beta}} \int_{\mathcal{C}_{\alpha}}P(\alpha, \beta)\frac{|\alpha\rangle \langle \beta^*|}{\langle \beta^*|\alpha\rangle}\mathrm{d}\alpha \mathrm{d}\beta,
\end{equation}
where $\mathcal{C}_{\beta}$ and $\mathcal{C}_{\alpha}$ are paths of integration on individual complex planes $(\alpha,\beta)$. This representation allows a mapping of the master equation into a Fokker-Planck equation for the function $P(\alpha, \beta)$. As a result, all  one-time correlation functions in the steady-state can be computed exactly and are given by:
\begin{align}
\langle (b^{\dagger})^j(b)^j \rangle =& \left|\frac{2F}{U}\right|^{2j}\times\frac{\Gamma(c)\Gamma(c^*)}{\Gamma(c+j)\Gamma(c^*+j)} \nonumber \\ 
&\times\frac{\mathcal{F}(j+c,j+c^{*},8|F/U|^2)}{\mathcal{F}(c,c^{*},8|F/U|^2)}, \label{expect}
\end{align}
with
$
c = 2(-\Delta \omega -i\gamma/2)/U
$
and the hypergeometric function $
\mathcal{F}(c,d,z) = \sum_n^{\infty} \frac{\Gamma (c) \Gamma (d)}{\Gamma(c+n) \Gamma(d+n)}\frac{z^{n}}{n!}
$,
 $\Gamma$ being the gamma special function. These results for the single cavity can be directly applied to mean-field theory by performing the substitution of Eq.(\ref{MF}) and solving the following self-consistent equation for $\langle b \rangle$:
 \begin{equation}
\langle b \rangle = \frac{(F-J\langle b \rangle)}{\Delta \omega+i\gamma/2}\times\frac{\mathcal{F}(1+c,c^{*},8|\frac{F-J\langle b \rangle}{U}|^2)}{\mathcal{F}(c,c^{*},8|\frac{F-J\langle b \rangle}{U}|^2)}. \label{self-consistent}
\end{equation}
 
In order to characterize the state of the intracavity electromagnetic field, we will focus mainly on two observables: the mean photon density $\langle b^{\dagger}b \rangle$, and the zero-delay second order autocorrelation function $g^{(2)}(0)$. The latter is expressed by: 
\begin{equation}
g^{(2)}(0) = \frac{\langle b^{\dagger}b^{\dagger}bb \rangle}{\langle b^{\dagger}b \rangle^2},
\end{equation}
and gives information on the photon statistics.  It was shown that the self-consistent parameter in Eq.(\ref{H_MF}) is responsible for the appearance of tunneling-induced bistability. That is, for a wide range of parameters, the self-consistent equation for $\langle b \rangle$ has multiple solutions, giving rise to two stable steady-states (for the same values of all the parameters). In particular, we have identified a `low-density' phase, where the average number of photons per site is very low ($\langle b^{\dagger}b \rangle \ll 1$) but fluctuations in the photon statistics are high ($g^{(2)}(0) \gg 1$). In the other stable phase, called `high-density' phase, $\langle b^{\dagger}b \rangle \sim 1$ and the statistics is sub-poissonian ($g^{(2)}(0)) < 1$). Interestingly, a related bistable behavior has been predicted for systems of driven-dissipative Rydberg atoms \cite{Rydberg}.

As mentioned in the introduction, the next question that comes to mind is how to connect these results with the better known physics of the equilibrium model and its quantum phase transition. Are the stationary states related in any way to a Mott insulator or a superfluid? The best way to answer this question is to explore the limit of weak pumping and weak dissipation, since in this regime the Hamiltonian and the dynamics tend to resemble more and more their equilibrium counterpart. It is the subject of the following section.
%
%
\section{ The limit of weak pumping and weak excitation}
\label{sec:WPE}
\subsection{Single Cavity: Exact Solution}
\label{subsec:single_cav}

{\it Multiphotonic resonances}. At equilibrium, the ground state of an isolated site is a pure Fock state $|n\rangle$, with $n$ fixed by the value of the chemical potential. Photons, on the other hand, have always to be injected inside the cavity. When the coupling to the external field is very weak, the only way to have $\langle b^{\dagger}b \rangle \geq 1$ is to be at resonance  with multiphotonic absorption processes.

Absorption of photons is favored when $n$ incident laser photons have the same energy than $n$ photons inside the cavity, that is: $n\omega_p = n\omega_c +Un(n-1)/2$. Expressed in terms of the pump-cavity detuning, this relation reads:
\begin{equation}
\frac{U}{\Delta \omega} = \frac{2}{n-1}.
\end{equation}
This resonance condition can be satisfied for $n > 1$ only if $\Delta \omega > 0$. If the pump is resonant with the bare cavity frequency, i.e $\Delta \omega = 0$,  only single photons can be absorbed resonantly. There is no relation between $U$ and $\Delta \omega$ in this case and the density matrix is found by expanding the master equation in powers of $F/U$ and $\gamma/U$. In the following, we will focus on the more interesting case of multiphotonic resonances ($n>1$) and assume $\Delta \omega > 0$. 

For simplicity, let us first look at $H_{mf}$ for $J = 0$ and $U = 2\Delta \omega$ (two-photon resonance). It appears that in the rotating frame, the vacuum $|0\rangle$ has the same energy that the two photon-state $|2\rangle$ in the absence of driving. This degeneracy is lifted by the coupling to the external field whose effect on the dynamics can be understood qualitatively in the following way. Suppose that at time $t = 0$ the intra-cavity field is in the vacuum state $|0\rangle$. Since the vacuum is no longer an eigenstate of the Hamiltonian, but a linear combination of eigenstates $\frac{1}{\sqrt{2}}(|0 \rangle+|2\rangle)$ and $\frac{1}{\sqrt{2}}(|0 \rangle-|2\rangle)$, the cavity field will start to oscillate between $|0\rangle$ and $|2\rangle$. These Rabi oscillations will take place until the occurrence of a quantum jump, resulting from spontaneous emission processes inside the cavity.  If the frequency splitting between the two eigenstates is very small when compared with the dissipation rate, the field will not have time to oscillate and will stay mostly in the vacuum state. However, the photons being still absorbed by pairs, the fluctuations in the photon distribution may be very high. 
On the contrary,  Rabi oscillation will take place if the dissipation rate is sufficiently small. Besides, the time between quantum jumps will be longer, thereby reducing the probability of emitting two photons at once.

\begin{figure}
	\includegraphics[width = 0.49\columnwidth]{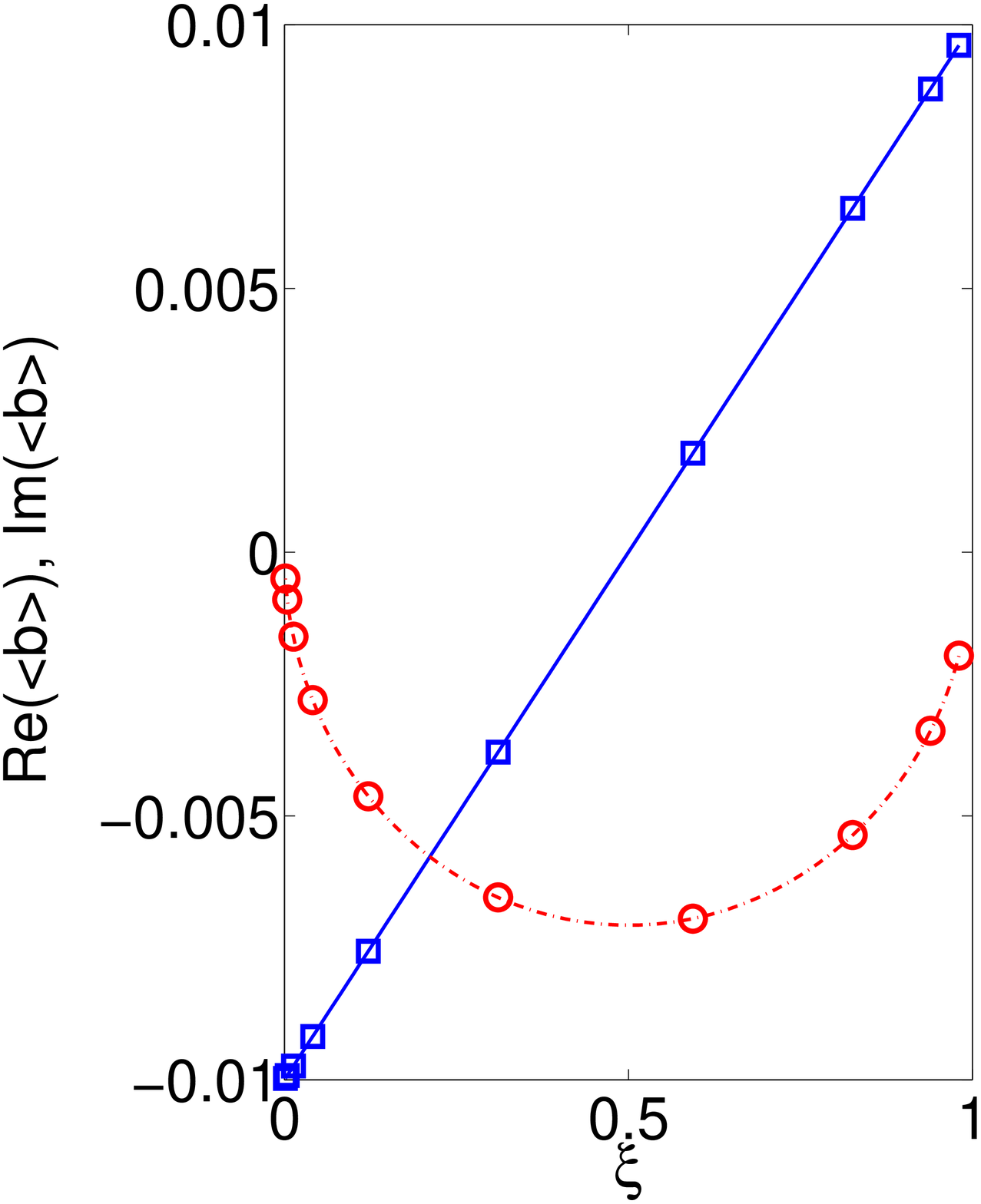}
	\includegraphics[width = 0.49\columnwidth]{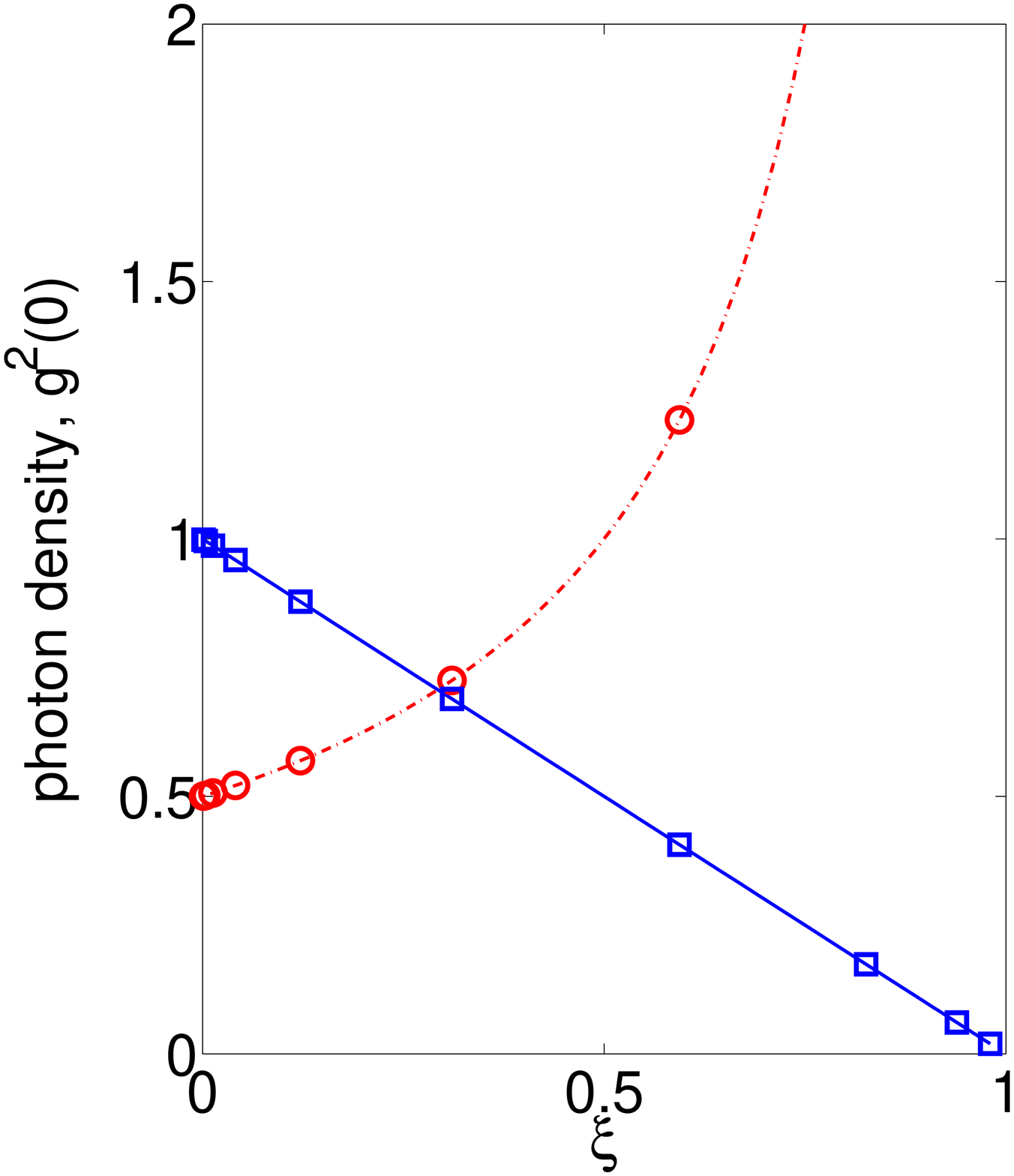}
\caption{(Color online) Results for a single cavity. Left panel: real (continuous blue line) and imaginary part (dotted-dashed red line), of the bosonic coherence $\langle b \rangle $ plotted vs $\xi$ for $F/\Delta \omega = 10^{-2}$. The corresponding values of $\gamma/\Delta \omega$ range from $\frac{F^2}{10\Delta \omega^2}$ to $\frac{F}{5\Delta \omega}$. Lines are the result of exact $P$-representation calculations while markers correspond to the simplified expression of Eq.(\ref{b_res}). Right panel: mean photon density $\langle b^{\dagger}b\rangle$ (continuous blue line) and $g^{(2)}(0)$ (red dashed-dotted line), vs. $\xi$. Same conditions and conventions as in the other panel.}
\label{fig:b_densg2}
\end{figure}

More quantitative results can be obtained by solving the master equation explicitly. To fully grasp the effect of the resonance,  we will assume that $F/\Delta \omega \gg \gamma/\Delta \omega$. 
We will work is the basis formed by the eigenstates of the total Hamiltonian (up to the lowest order in $F/\Delta \omega$), given by:
\begin{align}
|a\rangle & = \frac{1}{\sqrt{2}}(|0\rangle +|2\rangle),  \\
|b\rangle & = \frac{1}{\sqrt{2}}(|0\rangle - |2\rangle),  \\
|c\rangle & = |1\rangle. \nonumber
\end{align}  
Since the driving term only couples Fock states $|m\rangle$ to $|m \pm 1 \rangle$, the coupling of $|0\rangle$ and $|2\rangle$ is of order 2.  The energies given by second-order perturbation theory are:
\begin{align}
E_a &\simeq \frac{F^2}{\Delta \omega}(1+\sqrt{2}), \\
E_b & \simeq \frac{F^2}{\Delta \omega} (1-\sqrt{2}),    \\
E_c & \simeq -\Delta \omega.
\end{align}
As expected, the energy splitting between states $|a\rangle$ and $|b\rangle$ is proportional to $F^2/\Delta \omega$.
In the  ``dressed-states'' basis, the dissipative term of the master equation couples populations and coherences of the density matrix. However, in the lowest order in $F/\Delta \omega$ and $\gamma/\Delta \omega$, the coefficients $\rho_{ac}$ and $\rho_{bc}$ vanish. 
The master equation reads:
\begin{align}
\partial_t \rho_{aa} &= \gamma[\frac{1}{2}\rho_{cc}-\rho_{aa}+\frac{1}{2}(\rho_{ab}+\rho_{ba})],  \\
\partial_t \rho_{bb} &= \gamma[\frac{1}{2}\rho_{cc}-\rho_{bb}+\frac{1}{2}(\rho_{ab}+\rho_{ba})], \\
\partial_t \rho_{ab} &= \Delta \omega[(-i2\sqrt{2}\frac{F^2}{\Delta \omega^2}-\frac{\gamma}{\Delta \omega})\rho_{ab} +\frac{\gamma}{2\Delta \omega}] ,\\
\partial_t \rho_{cc} &= \gamma[\rho_{aa}+\rho_{bb}-\rho_{cc}-(\rho_{ab}+\rho_{ba})].  
\end{align}
The stationary value for $\rho_{ab}$ is then:
\begin{equation}
\rho_{ab} = \frac{1}{2}\frac{1}{1+\frac{i2\sqrt{2}F^2}{\gamma \Delta \omega}},
\end{equation}
giving:
\begin{equation}
\rho_{ab}+\rho_{ba} = \frac{1}{1+\frac{8F^4}{\gamma^2\Delta \omega^2}} = \xi.
\end{equation}
All the other coefficients along with the observables can be expressed as functions of the parameter $\xi$. Namely, we have:
\begin{align}
\rho_{11} &\simeq \rho_{cc} = \frac{1}{2}(1-\xi),\\
\rho_{aa} & = \rho_{bb} = \frac{1}{4}(1+\xi), \\
\rho_{22} &= \frac{1}{2}(\rho_{aa}+\rho_{bb})-\frac{1}{2}(\rho_{ab}+\rho_{ba}) = \frac{1}{4}(1-\xi).
\end{align}
For this we can compute the mean photon density and $g^{(2)}(0)$:
\begin{align}
\langle b^{\dagger }b \rangle &= 1-\xi  \label{dens_res}, \\
g^{(2)}(0) &= \frac{1}{2(1-\xi)} \label{g2_res}.
\end{align}
A comparison between theses two expressions and the exact $P$-representation formula is shown on Fig.(\ref{fig:b_densg2}). The mean photon density and $g^{(2)}(0)$ are plotted as a function of $\xi$ for $F/\Delta \omega = 10^{-2}$. In order to stay in the domain of validity of Eq.(\ref{dens_res}) and Eq.(\ref{g2_res}), $\gamma/\Delta \omega$ is ranging from $\frac{F^2}{10\Delta \omega^2}$ to $\frac{F}{5\Delta \omega}$. In this conditions, the approximations underlying the derivation are justified and the above expressions are very accurate.  

The two limits  $\frac{F^2}{\gamma \Delta \omega} \ll1$ and $\frac{F^2}{\gamma \Delta \omega}\gg 1$ correspond to $\xi \to 1 $ and $\xi \to 0$ respectively. In the first case, the photon density goes to zero as expected, and $g^{(2)}(0)$ diverges. In the other limit, the field is in a statistical mixture of 3 states and the density matrix in Fock space is:  
\begin{equation}
\rho = \frac{1}{4}|0\rangle \langle 0| + \frac{1}{2}|1\rangle \langle 1| +\frac{1}{4}|2\rangle \langle 2| \label{rho_2}.
\end{equation}

The photon density in then equal to one and $g^{(2)}(0) = 0.5$. We see that in the particular case of multiphotonic resonances, one must be careful in discussing the limit  $\frac{F}{\Delta \omega}, \frac{\gamma}{\Delta \omega} \ll 1$ as the system behavior varies qualitatively depending on the ratio $\frac{F^2}{\gamma \Delta \omega}$. 
The state obtained for $\xi \to 0$  is highly nonclassical and the closest to a Fock state that one can hope for in this context.

Besides, this result is not limited to $n = 2$.  In the general case of $n$-photon resonance, the coupling between $|0\rangle$ and $|n\rangle$ is of order $n$ in $F$ and the energy splitting proportional to $(F/\Delta \omega)^n$. Therefore, a state similar to Eq.(\ref{rho_2}) may be obtain in the limit $\frac{F^n}{\gamma \Delta \omega^{n-1}}\gg 1$. As shown in the Appendix, the corresponding density matrix is:
\begin{equation}
\rho^{(n)} = \frac{1}{2^{n}}\sum_{k=0}^n\begin{pmatrix}n\\k \end{pmatrix}|k\rangle\langle k| \label{rho_n}.
\end{equation}
For $n>2$, not only the two states $|0\rangle$ and $|n\rangle$ are degenerate in the absence of driving, but so are all the states $|k\rangle$, $|q\rangle$ with $k+q = n$. This degeneracy is reflected in the $n \to n-k$ symmetry of Eq.(\ref{rho_n}). The state of Eq.(\ref{rho_n}) is characterized by:
\begin{align}
\langle b^{\dagger }b \rangle &= \frac{n}{2},\label{dens_lim} \\
g^{(2)}(0) &= 1-\frac{1}{n}. \label{g2_lim}
\end{align}
Note that the value of $g^{(2)}(0)$ is the same as in the $n^{th}$-lobe of the equilibrium model (pure Fock state with $n$ photons). We emphasize that Eq.(\ref{rho_n}-\ref{g2_lim}) also apply to $n= 1$ ($\Delta \omega = 0$), in the limit $F/U\to 0$ and $\gamma/F \to 0$. 

%
%

{\it Driving out of multiphotonic resonances}. In the case where multiphotonic absorption processes are non resonant, the mean photon number is expected to be very low. By performing an expansion of the master equation in powers of $F/\Delta \omega$ and $\gamma/\Delta \omega$ one can show that the coefficients of the stationary density matrix (in Fock space) obey the following hierarchy: $Re(\rho_{n m}) \sim (\frac{F}{\Delta\omega})^{n+m}$ and $Im( \rho_{nm}) \sim \frac{\gamma}{\Delta \omega}(\frac{F}{\Delta\omega})^{n+m}$.  $F/\Delta \omega$ and $\gamma /\Delta \omega$ are assumed to be both much smaller than one, but it is not necessary to impose $F \gg \gamma $. The following treatment is still valid if $F/\Delta \omega$ and $\gamma /\Delta \omega$ are of the same order of magnitude. 

Neglecting the probability of having 3 or more photons inside the cavity and keeping the lowest order in $F/\Delta \omega$ and $\gamma /\Delta \omega$ for the remaining coefficients, we obtain the following set of equations (the real part of a complex number $z$ is denoted by $\overline z$, its imaginary part by $\tilde z$):   
\begin{align}
\overline \rho_{10} &= \frac{F}{\Delta \omega} \rho_{00},\\
\tilde \rho_{10} &= -\frac{\gamma}{2\Delta \omega}\overline \rho_{10},\\
\rho_{11} &= -\frac{2F}{\gamma}\tilde \rho_{10}, \\
\overline \rho_{20} &= \frac{\sqrt{2}F/\Delta \omega}{2-U/\Delta \omega}\overline \rho_{10},\\
\tilde \rho_{20} &= \frac{1}{2-U/\Delta \omega}(-\frac{\gamma}{\Delta \omega} \overline \rho_{20} + \frac{F}{\Delta \omega} \sqrt{2} \tilde \rho_{10}),\\
\overline \rho_{21} &= \frac{F/\Delta \omega}{1-U/\Delta \omega}(\sqrt{2} \rho_{11}-\overline \rho_{20}),\\
\tilde \rho_{21} &= \frac{-1}{1-U/\Delta \omega}(\frac{3\gamma}{2\Delta \omega}\overline \rho_{21}+\frac{F}{\Delta \omega} \tilde \rho_	{20} ),\\
\rho_{22} &= -\frac{\sqrt{2}F}{\gamma}\tilde \rho_{21}.
\end{align}
Setting $\epsilon = F/\Delta \omega$; $\eta = \gamma /\Delta \omega$ and $u = U/\Delta \omega$, the stationary density matrix reads:
\begin{equation}
\rho = 
\begin{pmatrix}
1 & \epsilon(1+\frac{i\eta}{2}) & \frac{\sqrt{2}\epsilon^2}{2-u}(1+\frac{i\eta}{2}\frac{4-u}{2-u}) \\
\epsilon(1-\frac{i\eta}{2}) & \epsilon^2 & \frac{\sqrt{2}\epsilon^3}{2-u}(1+\frac{i\eta}{2-u})\\
\frac{\sqrt{2}\epsilon^2}{2-u}(1-\frac{i\eta}{2}\frac{4-u}{2-u})&  \frac{\sqrt{2}\epsilon^3}{2-u}(1-\frac{i\eta}{2-u})& \frac{2\epsilon^4}{(2-u)^2}
\end{pmatrix},
\label{dens_mat}
\end{equation}
From this we can compute the mean photon density and the second-order autocorrelation function:
\begin{align}
\langle b^{\dagger}b\rangle &= \left(\frac{F}{\Delta \omega}\right)^2,\\
g^{(2)}(0) &= \frac{4}{(2-U/\Delta\omega)^2}.
\end{align} 
As expected, when the system becomes linear, i.e. $U \to 0$, the cavity is driven into a coherent state ($g^{(2)}(0) = 1$). However, the on-site interaction induces large fluctuations in the photon statistics when the two-photon absorption process becomes resonant ($U/\Delta \omega = 2$).
%
\subsection{Coupled Cavities: Mean-Field Solution}
\label{subsec:coupled_cav}

\begin{figure}
	\includegraphics[width = 0.49\columnwidth]{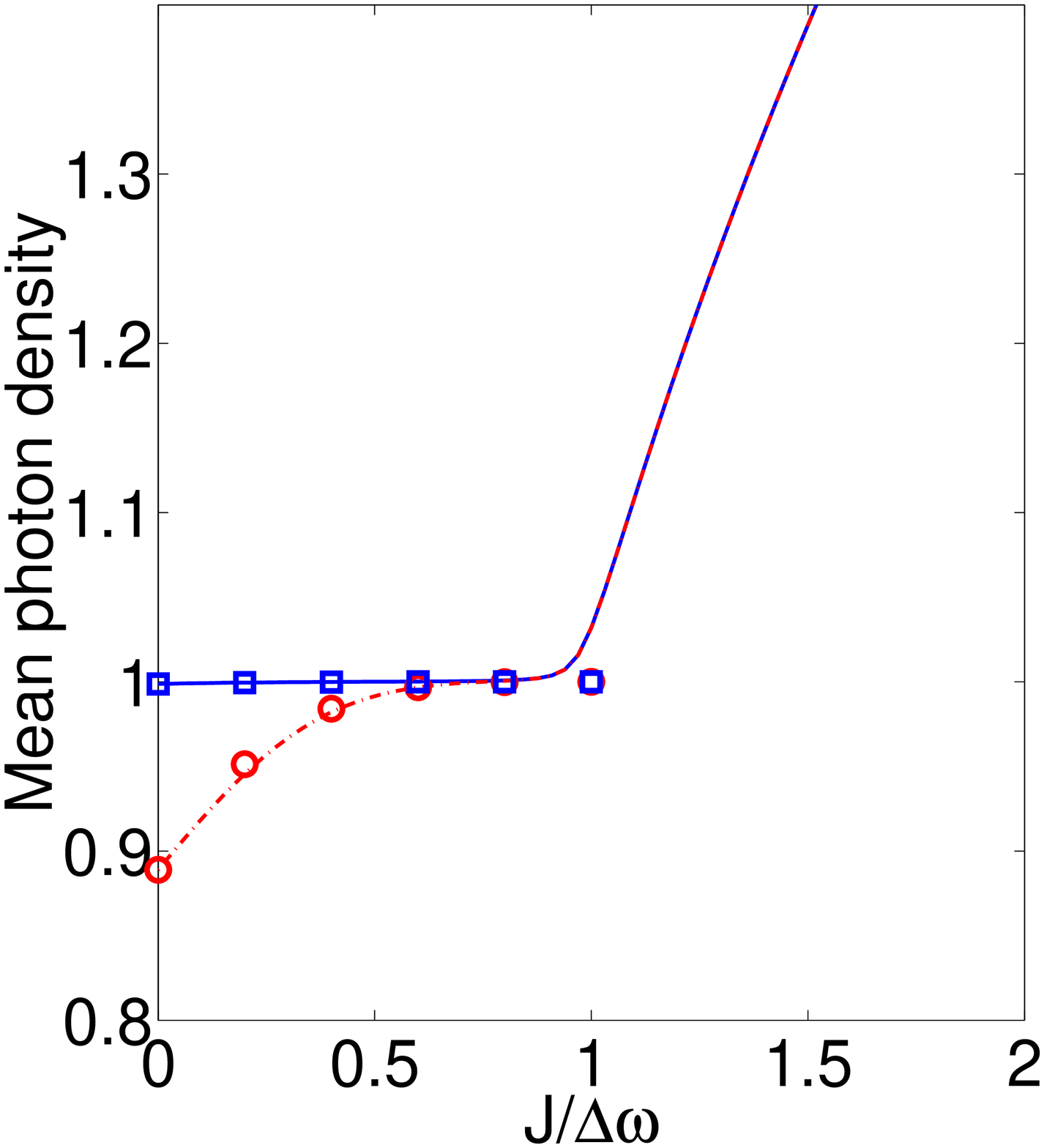}
	\includegraphics[width = 0.49\columnwidth]{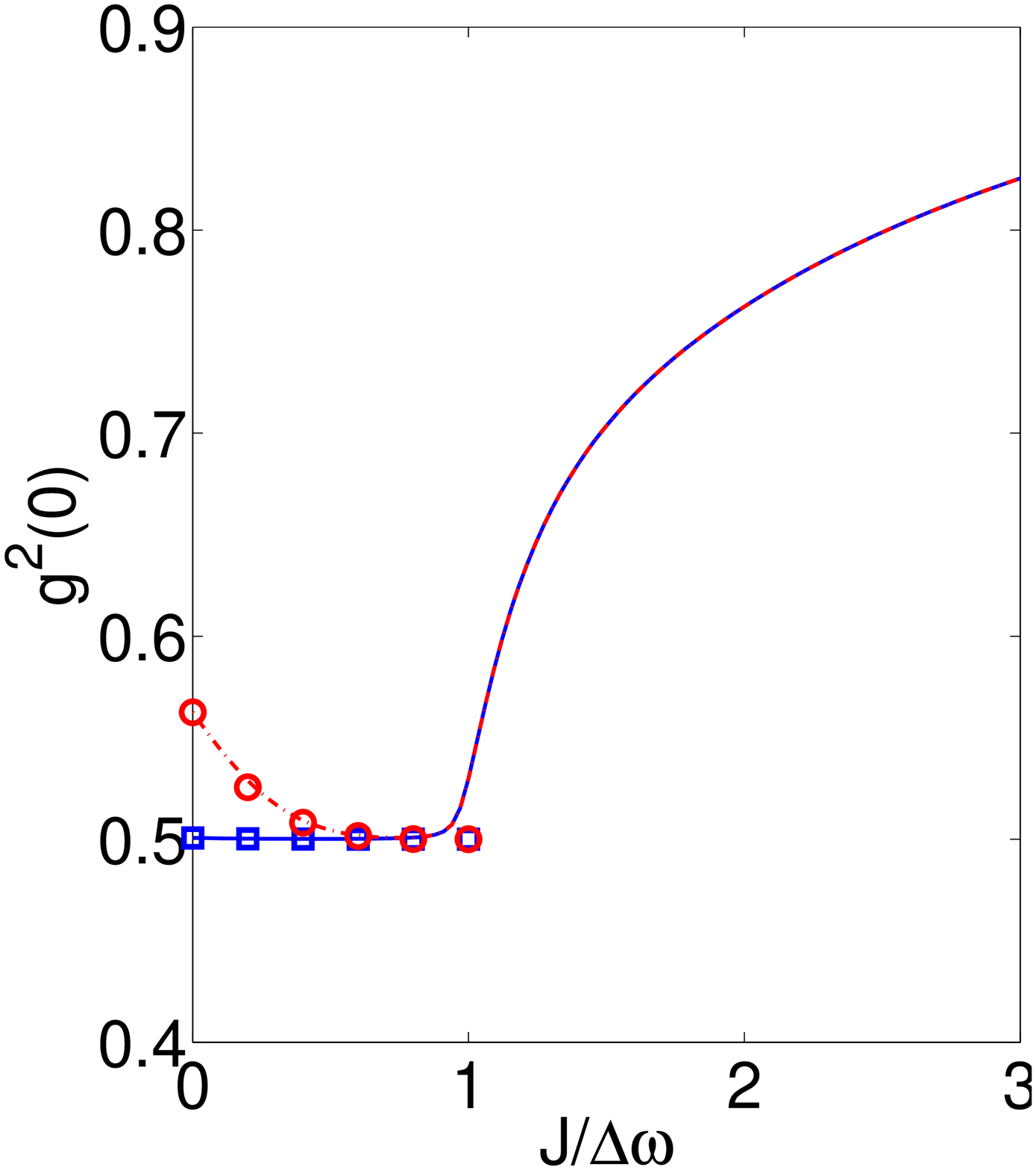}
\caption{(Color online) Two-photon resonance. Mean photon density and $g^{(2)}(0)$ as a function of the tunneling amplitude $J/\Delta \omega$ for $F/\Delta \omega = 10^{-2}$. The values of $\gamma/\Delta \omega$ are $\frac{F^2}{10\Delta \omega^2}$ (continuous blue line) and $\gamma/\Delta \omega = \frac{F^2}{\Delta \omega^2}$ (red dotted-dashed line). The lines show $P$-representation calculations and the markers the results of Eq.(\ref{b_MF_res}). The effect of coupling $J$ is to drive the system into the state of Eq.(\ref{rho_2}), until the critical coupling $J_c = \Delta \omega $ is reached.}
\label{fig:densg2_J}
\end{figure}

Let us first go back to the two-photon resonance. The analytical expression for the bosonic coherence in this regime is:
\begin{equation}
\langle b \rangle = \frac{F}{\Delta \omega}(2\xi-1)+i\frac{\gamma}{2F}(\xi-1) \label{b_res},
\end{equation}
and the mean-field self-consistent equation is obtained by replacing $F$ with $F-J\langle b \rangle $. Since $\xi$ is also a function of $F$, this equation is difficult to solve analytically in its general form. For $\xi \to 0$, however, $\gamma/F \ll 1$ and the imaginary part can be neglected. We are left with the simple expression:
\begin{equation}
\langle b \rangle = -\frac{F}{\Delta \omega}. \label{b_lim}
\end{equation}
 The substitution $F \to F-J\langle b \rangle$ then gives:
 \begin{equation}
\langle b \rangle = \frac{-F/\Delta \omega}{1-J/\Delta \omega} \label{b_MF_res}.
 \end{equation} 
This shows that the coupling between sites amounts to replacing $F$ with $F'  = \frac{F}{1-J/\Delta \omega}$. In other words, the effective pump is enhanced by the coupling between cavities. As a result, the system is driven into the $\xi = 0$ state and will stay there as long as the approximation $F'/\Delta \omega \ll 1$ holds. Results for different values of $\xi$ are presented on Fig.(\ref{fig:densg2_J}). When $J/\Delta \omega \sim 1$ the above treatment ceases to be valid because $F' \sim \Delta \omega$, and the system enters another regime.  Exact $P$-representation calculations show that the mean photon density starts to increase with $J$, while $g^{(2)}(0)$ goes to 1, thus indicating a crossover from a quantum state to a classical coherent one (see Fig.(\ref{fig:densg2_J})).  As we shall see in section $\ref{sec:GP}$, this idea is confirmed by the fact that the linear asymptotic behavior of $\langle b^{\dagger}b \rangle$ as a function of $J$ visible on Fig.(\ref{fig:densg2_J}), corresponds to Gross-Pitaevski semi-classical predictions.

\begin{figure}
	\includegraphics[width = 0.49\columnwidth]{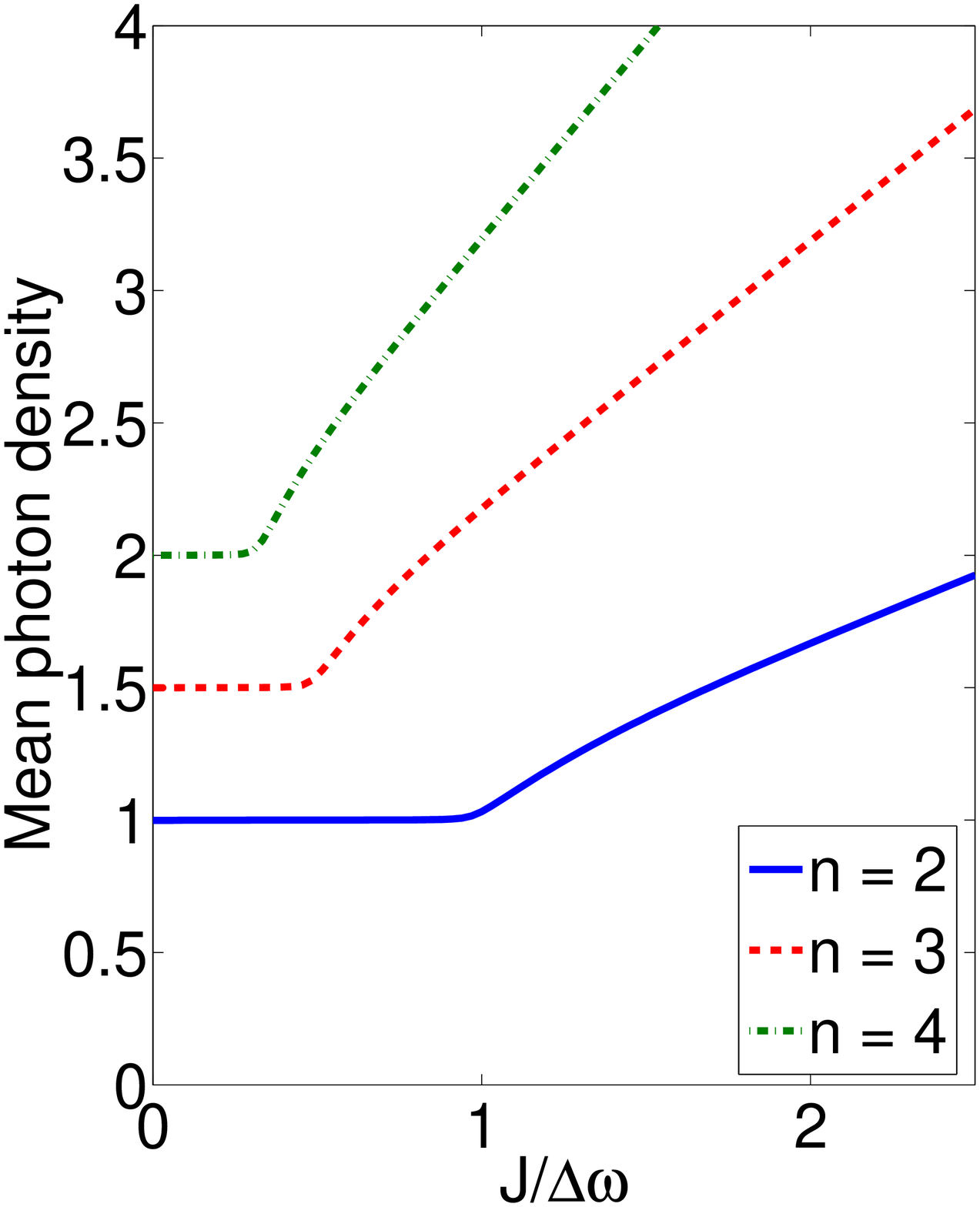}
	\includegraphics[width = 0.49\columnwidth]{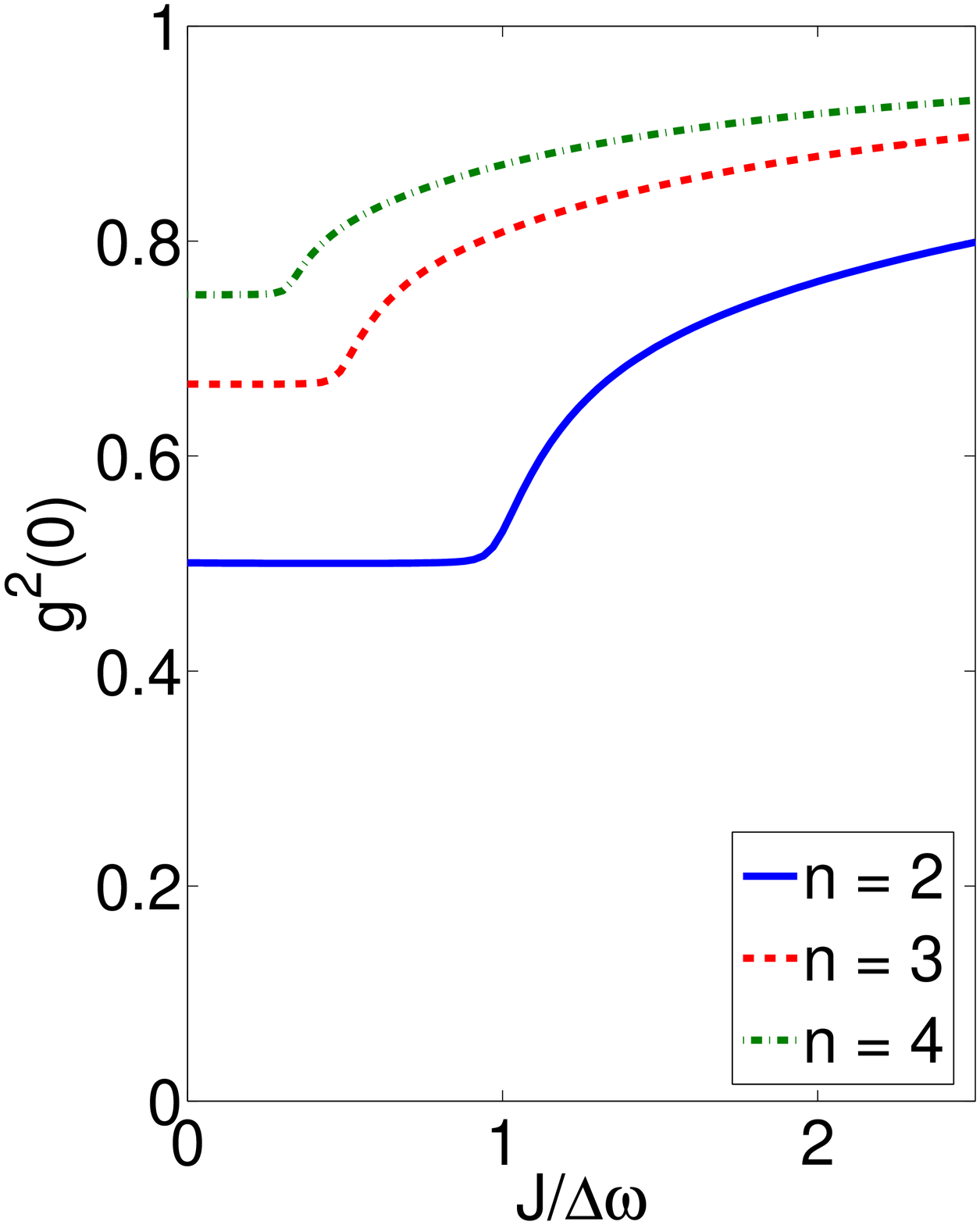}
\caption{(Color online) Quantum to classical crossover for $n = 2$ (continuous blue line), $n = 3$ (red dashed line), and $n = 4$ (green dotted-dashed line). The mean photon density and $g^{(2)}(0)$ are plotted as a function of the tunneling amplitude $J/\Delta \omega$ for $F/\Delta \omega = 10^{-2}$ and $F^n/(\Delta \omega ^{n-1}\gamma) = 10$. The system stays in the state of Eq.(\ref{rho_n}) until a critical coupling $J_c = \Delta \omega/(n-1)$, after which it is driven towards a coherent state.}
\label{fig:n_phot}
\end{figure}

Once again, this can be extended to larger values of $n$. For the $n$-photon resonance in the limit $\frac{F^n}{\gamma \Delta \omega^{n-1}}\gg 1$, Eq.(\ref{b_lim}) for a single cavity becomes (see Appendix):
\begin{equation}
\langle b \rangle = -(n-1)\frac{F}{\Delta \omega}. \label{b_lim_n}
\end{equation}
When the coupling between cavities in switched on, $\langle b \rangle$ is given by:
\begin{equation}
\langle b \rangle = \frac{-(n-1)F/\Delta \omega}{1-(n-1)J/\Delta \omega}, \label{b_MF_res_n}
\end{equation}
which means that the system will stay in the state Eq.(\ref{rho_n}) until $J/\Delta \omega \sim 1/(n-1)$. As in the case of two-photon resonance, exact $P$-representation calculations presented on Fig.(\ref{fig:n_phot})  show a crossover to a classical coherent state. This crossover is the closest equivalent, in this driven dissipative system, of the equilibrium Mott insulator to superfluid phase transition.

 
Out of multiphotonic resonances, the coupling between cavities has a different effect. In this regime, the system is in a state described by Eq.(\ref{dens_mat}) at $J = 0$. The bosonic coherence is then:
\begin{equation}
\langle b \rangle  = \frac{F}{\Delta \omega}.
\end{equation} 
At finite $J$, it becomes:
\begin{equation}
\langle b \rangle = \frac{F/\Delta \omega}{1+J/\Delta \omega}, \label{b_MF_nonres}
\end{equation}
and the effective pump is given by $F  \to F' = \frac{F}{1+J/\Delta \omega}$. Contrary to Eq.(\ref{b_MF_res_n}), the intensity of $F'$ decreases with $J$. As a consequence, the system will remain in a state qualitatively similar to Eq.(\ref{dens_mat}) and no crossover occurs.

\subsection{Relation to Steady-States at Higher Pumping and Dissipation}

The results presented above shed light on the nature of the steady-states obtained at higher pumping and dissipation. First, the effect of multiphotonic resonances is visible on the bistability diagram: they are responsible for its peculiar lobe structure (see Fig.(\ref{fig:bistab}) or ref. \cite{PRL2013}). Moreover, the properties of the states of Eq.(\ref{rho_n}) and the behavior showed on Fig.(\ref{fig:densg2_J}) and Fig.(\ref{fig:n_phot}) is very similar to that of the `high-density phase mentioned in section \ref{sec:model}. In this phase the light is antibunched and the photon density increases with increasing  $J$.

As for the `low-density' phase, it shows photon superbunching, near the two-photon resonance, which is well described by Eq.(\ref{dens_mat}). Besides, the photon density in this phase is decreasing with $J$, as suggested by Eq.(\ref{b_MF_nonres}).

Comparing the expression for the bosonic coherence in Eq.(\ref{b_MF_nonres}) and Eq.(\ref{b_MF_res_n}) we see that in both regimes, the effect of tunneling  is directly related to the sign of its real part. Interestingly, this remains true for higher pumping and dissipation, although the mean-field self-consistent equation takes a more complicated form. The real part of $\langle b \rangle $ is negative in the `high-density' phase and positive in the `low-density' phase.
 
%
\section{The Gross-Pitaevski Regime}
\label{sec:GP}

\begin{figure}
\begin{center}
\includegraphics[width = \columnwidth]{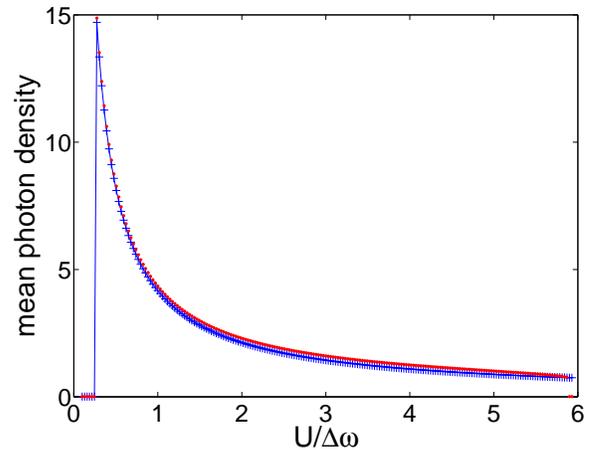}
\end{center}
\caption{(Color online) Mean photon density versus  on-site repulsion $U$ in the high density phase, for $J/\Delta \omega = 3$ ; $F/\Delta \omega =0.4$ ; $\gamma/\Delta \omega = 0.2$. Red dots: $P$-representation calculations ; Blue crosses: Gross-Pitaevski approximation.} 
\label{fig:dens_max}
\end{figure}

We have seen in section \ref{sec:WPE}, in the case of the multiphotonic resonances, that as $J/\Delta \omega$ becomes large, the system is driven into an almost coherent steady-state with $b^{\dagger}b \gg1$. This indicates that it enters a semi-classical regime where correlation functions can be approximated by :
\begin{equation}
\langle b^{\dagger n} b^{m} \rangle \simeq \langle b^{\dagger} \rangle^{ n} \langle b \rangle ^m. \label{GP_approx}
\end{equation}
As a consequence, all these functions are determined by a single complex number, namely the bosonic coherence $ \beta = \langle b \rangle $.  Besides, a general differential equation for correlation functions can be readily obtained from Eq.(\ref{ME}).  Its most general expression in the context of mean-field therory is the following:
\begin{equation}
\partial_t \langle b^{\dagger n}b^m \rangle = \langle [b^{\dagger n}b^m,H_{mf}] \rangle  -\frac{i\gamma}{2}(n+m)\langle b^{\dagger n}b^m \rangle. \label{Eqdiff_gen}
\end{equation}
In the particular case of $\beta = \langle b \rangle$ and under the assumption of Eq.(\ref{GP_approx}), the previous equation yields:
\begin{equation}
i\partial_{t} \beta = (-\Delta \omega-J- \frac{i\gamma}{2} +U|\beta|^2 )\beta+F. \label{GP}
\end{equation}
 This equation is a single-mode version of the Gross-Pitaevski equation. Note that in this regime, the decoupling of neighboring sites amounts to a shift in the cavity frequency, $\Delta \omega \to \Delta \omega+J $. The steady-state value for $\beta$ is:
 \begin{equation}
 \beta = \frac{F}{\Delta \omega +J-U|\beta|^2+\frac{i\gamma}{2}},
 \end{equation}
 which gives a third order polynomial equation for the mean photon density $n = |\beta|^2$:
 \begin{equation}
 n((\Delta \omega+J-nU)^2+\frac{\gamma^2}{4}) = F^2. \label{Eq_dens}
 \end{equation}
 This equation explains the linear assyptotic behavior of $n$ as a function of $J/\Delta \omega$ visible on Fig.(\ref{fig:densg2_J}). Indeed, when $F,\gamma, \to 0$ and $J \to \infty$, we find:
 \begin{equation}
 n  \sim \frac{J}{U},
 \end{equation}
 which agrees with the results of Fig.(\ref{fig:densg2_J}) and Fig.(\ref{fig:n_phot}).
 Gross-Pitaevski approximation is also relevant at higher pumping and dissipation, especially when the coupling between sites and the number of photons are very high. As it as been widely use in the theory of quantum fluids, whether with cold atoms or polaritons, it is fruitful to compare Gross-Pitaevski results with $P$-representation calculations presented in \cite{PRL2013}.  For example, Fig.(\ref{fig:dens_max}) shows  that for large coupling between sites and in the `high density' phase, Gross-Pitaevski approximation is sufficient to capture the behavior of the mean photon density as a function of the on-site interaction $U$.
 
\subsection{Gross-Pitaevski Criterium for Bistability}

Eq.(\ref{Eq_dens}) was introduced in quantum optics as part of a semi-classical theory of optical bistability in a single nonlinear cavity \cite{DW}. In the present context, as mentioned in section \ref{sec:model}, mean-field theory predicts tunneling-induced bistability within a fully quantum framework.
For some values of the parameters, Eq.(\ref{Eq_dens}) has three real and positive roots. One of them corresponds to the `low-density' phase ($n \sim 10^{-2}$) and the two others to `high-density phases'. Although only one `high-density' phase was mentioned in our previous description of the mean-field phase diagram, a third solution was indeed found using generalized $P$-representation, but the corresponding phase proved to be always unstable. 

\begin{figure}
	\includegraphics[width = \columnwidth]{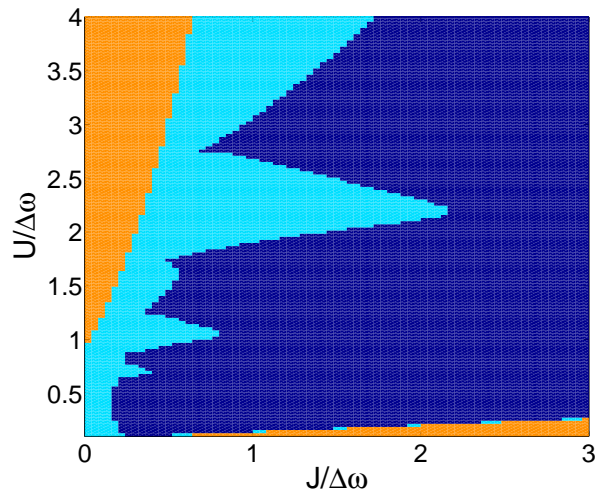}
	\caption{(Color online) Gross-Pitaevski and $P$-representation bistability diagrams. Orange: monostable phase according to both approximation schemes ; light blue: bistable according to Eq.(\ref{Eq_dens}) but monostable according to $P$-representation calculations ; dark blue: bistable phase according to both approximation schemes.}
\label{fig:bistab}
\end{figure}

Fig.(\ref{fig:bistab}) shows the two bistability diagrams obtained respectively from Eq.(\ref{Eq_dens}) and  generalized $P$-representation. As expected, Gross-Pitaevski approximation is very good for small values of $U$,  and predicts accurately the appearance of bistability in the lower-right corner of the diagram. It is less accurate when $U$ becomes large and on the whole, bistability is ``overestimated'' by the Gross-Pitaevski criterium: monostable regions according to Eq.(\ref{Eq_dens}) (in orange on Fig.(\ref{fig:bistab})), are much smaller that the exact ones (in light blue). In particular it fails to predict the lobe structure that is visible on the $P$-representation diagram. These lobes stem from the $n$-photon resonances discussed in the previous section.  Since a semi-classical approach does not take into account the quantized nature of the field, these resonances are washed-out in the Gross-Pitaevski diagram. 
%

%
%
In the framework of Gross-Pitaevski approximation, the number of solutions is given by the sign of the discriminant of Eq.(\ref{Eq_dens}). A very good approximation for the critical value of $U$ can be found by noticing that in the high-density phase, the photon density decreases with $U$. The critical value is then approximately the one for which the density is maximal. This yields:
\begin{equation}
\frac{U_{c1}}{\Delta \omega} = \frac{\gamma^2}{4F^2}(1+\frac{J}{\Delta \omega}). \label{Eq_edge}
\end{equation}
In fact, this approximate expression corresponds to the first term in the expansion in powers of $F/\Delta \omega$ of the exact solution. A similar expansion for the other frontier in the diagram gives:
\begin{equation}
\frac{U_{c2}}{\Delta \omega} = \frac{4\Delta \omega^2}{27F^2}(1+J/\Delta \omega )^3.
\end{equation}

The expansion up to the next term is:   
\begin{align}
\frac{U_{c1}}{\Delta \omega} &= \frac{\gamma^2}{4F^2}(1+\frac{J}{\Delta \omega}) -\frac{\gamma^4}{64\Delta \omega^2 F^2(1+J/\Delta \omega)},\\
\frac{U_{c2}}{\Delta \omega} &= \frac{4\Delta \omega^2}{27F^2}(1+J/\Delta \omega )^3 +\frac{\gamma^2}{12F^2}(1+J/\Delta \omega).
\end{align} 

\subsection{Bogoliubov Theory}

 As shown in our previous work, one can study fluctuations of the density matrix around mean-field by means of a extended  Bogoliubov theory. The fluctuations are defined as follow: 
\begin{equation}
\rho = \bigotimes_{i}(\rho_{mf}+\delta \rho_i).
\end{equation}
We also introduce  the Fourier transform of the matrices $\delta \rho_{i}$, $ \delta \rho_{\bf k } = \frac{1}{\sqrt{N}}\sum_{i = 1}^{N} e^{-i{\bf k  \cdot r_i}}\delta \rho_{i}$\\

In their most general formulations, the equations of evolution that stem from linearization around mean-field are:
\begin{equation}
i\partial_{t} \delta \rho_{\bf k} =  {\mathcal L}_{mf}[\delta \rho_{\bf k}] + {\mathcal L}_{\bf k}[\delta \rho_{\bf k}], \label{Bogo}
\end{equation}
where 
\begin{equation}
{\mathcal L}_{mf}[\delta \rho_{\bf k}] = [H_{mf}, \delta \rho_{\bf k}] - \frac{i\gamma}{2}(2b\delta \rho_{\bf k}b^{\dagger}-b^{\dagger}b\delta \rho_{\bf k}-\delta \rho_{\bf k}b^{\dagger}b).
\end{equation}
This operator is the usual Liouvillian for the effective single cavity problem. This term in Eq.(\ref{Bogo}) is thus independent of ${\bf k}$. Propagation effects arise from the second term:
\begin{equation}
{\mathcal L}_{\bf k}[\delta \rho_{\bf k}] = -t_{\bf k}({\rm Tr}(b \delta \rho_{\bf k})[b^{\dagger}, \rho_{mf}]+{\rm Tr}(b^{\dagger} \delta \rho_{\bf k})[b, \rho_{mf}]),
\end{equation} 
with $t_{\bf k} = J/z(\cos k_{x}a+\cos k_ya)$.

 The situation is greatly simplified in the Gross-Pitaevski regime where the system is described by classical complex field. Fluctuations around the mean-field value $\beta$ then obey the following equation:
\begin{widetext}
\begin{equation}
i\partial_t
\begin{pmatrix}
\delta \beta_{\bf k}\\
\delta \beta^{*}_{\bf -k}
\end{pmatrix}
 = 
 \begin{pmatrix}
 -\Delta \omega -t_{\bf k}+2U|\beta|^2-i\gamma/2 & U\beta ^2\\
 U \beta^{*2} & \Delta \omega +t_{\bf k}-2U|\beta|^2-i\gamma/2 
 \end{pmatrix}
 \begin{pmatrix}
\delta \beta_{\bf k}\\
\delta \beta^{*}_{\bf -k}
\end{pmatrix}.
\end{equation}
\end{widetext}

This leads to a complex Bogoliubov spectrum:
\begin{equation}
\omega_{\pm}({\bf k}) = \pm \sqrt{(-\Delta \omega-t_{k}+2U|\beta|^2)^2-U^2|\beta|^4}-\frac{i\gamma}{2}. \label{Bogo_GP}
\end{equation}
\begin{figure}
	\includegraphics[width = 0.49\columnwidth]{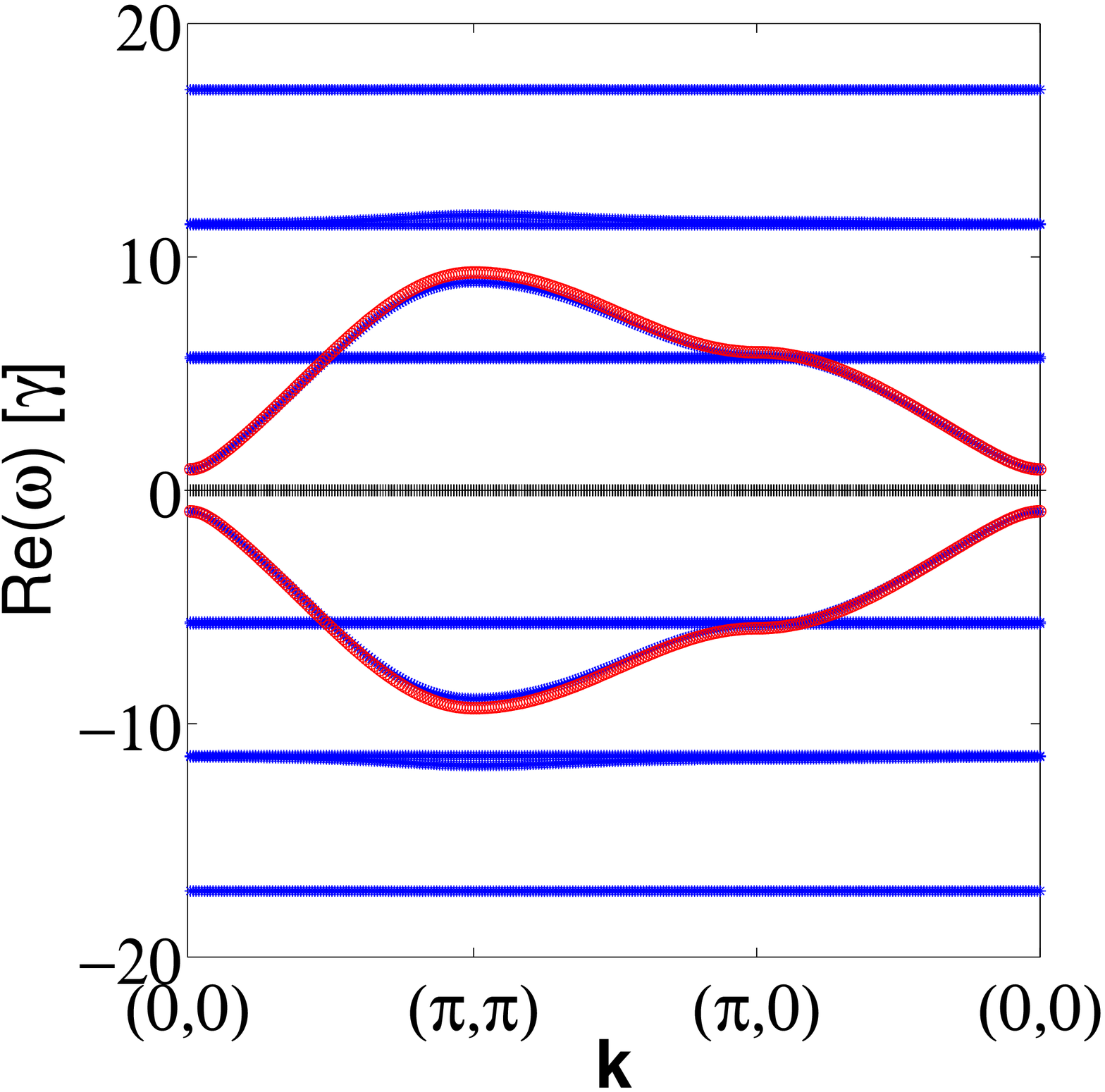}
	\includegraphics[width = 0.49\columnwidth]{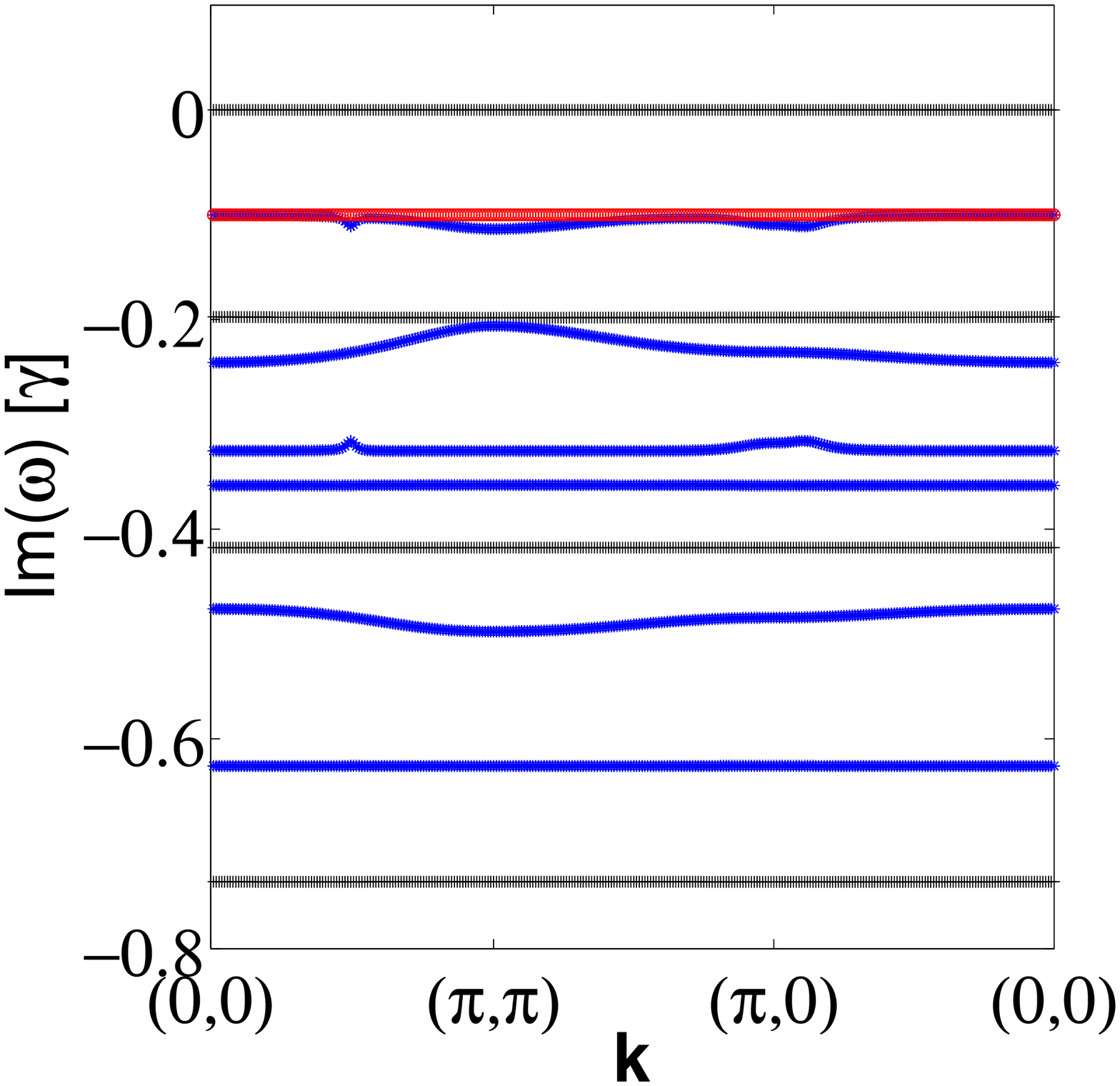}\\
	\includegraphics[width = 0.49\columnwidth]{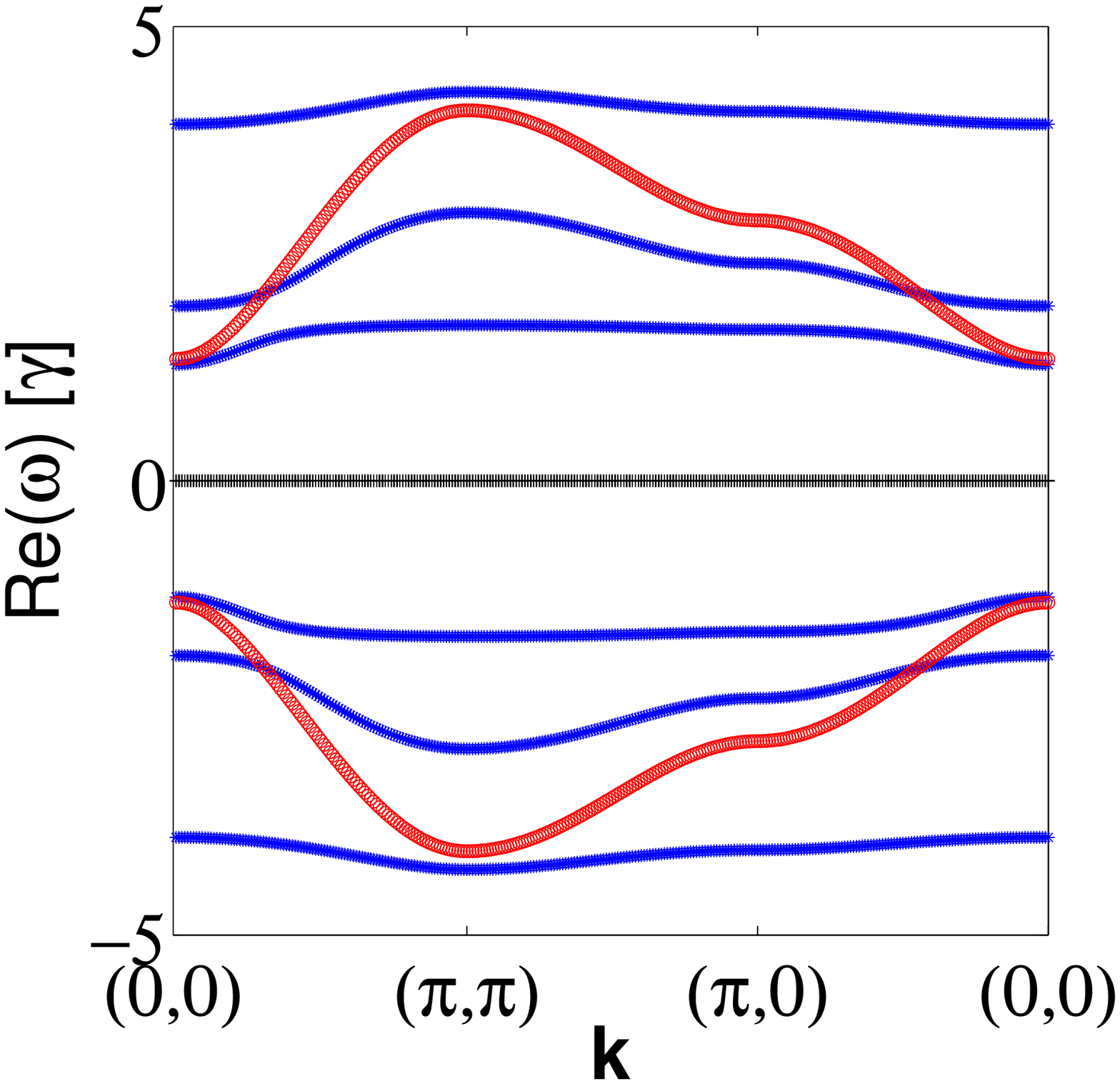}
	\includegraphics[width = 0.49\columnwidth]{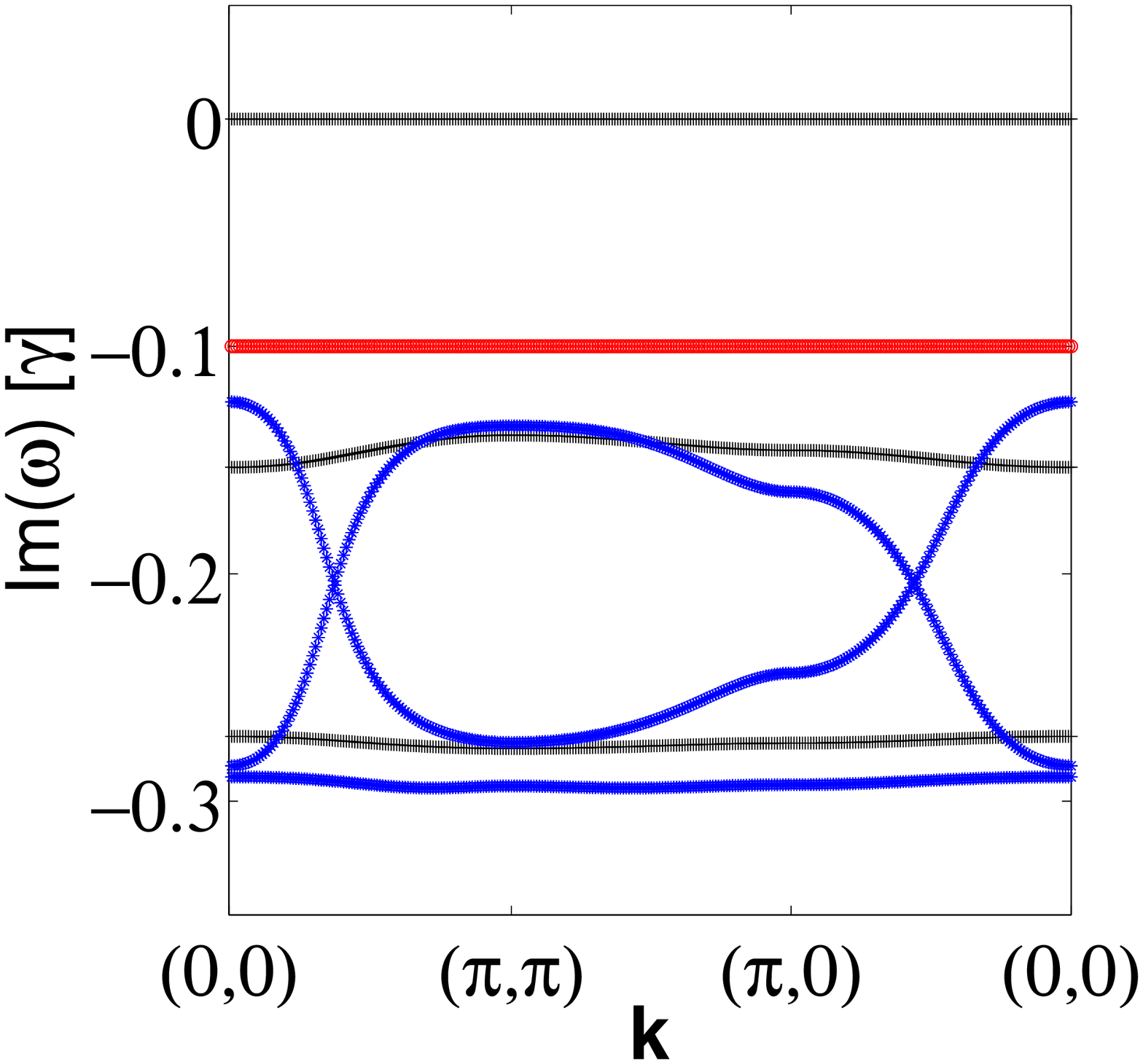}
\caption{(Color online) Energy-momentum dispersion of elementary excitations. Upper panels:  $\gamma/\Delta \omega = 0.2$ ; $F/\Delta \omega = 0.4$, $U/\Delta \omega = 0.5$ and $J/\Delta \omega = 3 $ (high-density phase). Real and imaginary part of the low-energy branches (in units of $\gamma$)  are plotted vs $\mathbf{k}$. Blue and black lines depict branches obtained with Eq.(\ref{Bogo}), the two red lines are the branches derived from Gross-Pitaevski equations. For these parameters, Gross-Pitaevski approximation is accurate.
Lower panels: $\gamma/\Delta \omega = 0.2$ ; $F/\Delta \omega = 0.4$, $U/\Delta \omega = 2$ and $J/\Delta \omega = 1 $ (monostable phase). Gross-Pitaevski approximation fails in the regime of strong correlations.}
\label{fig:disp}
\end{figure} 
Dispersion relations extracted from Eq.(\ref{Bogo}) and Eq.(\ref{Bogo_GP}) are shown on Fig.(\ref{fig:disp}). For small on-site repulsion and large tunneling amplitude ( $U/\Delta \omega = 0.5$ and $J/\Delta \omega = 3 $, upper panels), the Gross-Pitaevski approximation give good quantitative results and the corresponding spectrum is included in the more general approached outlined in Eq.(\ref{Bogo}). As expected, it fails in the regime of strong correlations. Lower panels of Fig.(\ref{fig:disp}) show the dispersion relations for $U/\Delta \omega = 2$ and $J/\Delta \omega = 1$. For these parameters, $g^{(2)}(0) = 0.69$, proving that the hypothesis of a quasi coherent state underlying Gross-Pitaevski approximation scheme is not justified.  

%
%
\section{Conclusion}
\label{sec:conclu}
In this paper we have explored the driven-dissipative Bose-Hubbard model in the limit of weak pumping and weak dissipation and provided analytical results for the mean-field density matrix. In this regime, the driven-dissipative model bear a formal resemblance to its equilibrium counterpart. However, since photons, unlike, atoms have to be injected by the pump laser, the mean photon number inside the cavity can be large only when multiphotonic absorption processes become resonant. In the case of $n$-photon resonance we have shown that if the intensity of the pump is sufficiently large when compared with the dissipation rate, a single cavity can be driven into a statistical mixture that has the same second-order correlation function $g^{(2)}(0)$ as a pure Fock state with $n$ photons, and a mean photon number of $n/2$. At resonance, the effect of the coherent pump is enhanced by the coupling between sites, eventually leading to a crossover from these quantum states to classical coherent ones.This behavior is characteristic of the `high-density' phase observed at higher pumping and dissipation, in the regime of tunneling-induced bistability.  Outside of these multiphotonic resonance processes, the mean-photon density is much smaller than 1 and the effect of the pump is reduced by tunneling. Nevertheless, on-site interactions induce photon superbunching close to the two-photon resonance. This peculiar photon statistics is recovered at higher pumping and dissipation in the `low-density' phase of the bistable region.    

In addition, we have shown that the structure of the bistability diagram cannot be explained without a full quantum treatment of the single cavity. Indeed, a Gross-Piaevski semiclassical approach gives satisfactory results in the `weakly-interacting' sector of the diagram, but fails in the strongly correlated regime. In particular, the size of the bistable region predicted by Gross-Pitaevski equation is considerably larger when compared with P-representation mean-field calculations.

\section{ACKNOWLEDGEMENT}

We acknowledge support by the French ANR project QUANDYDE and by ERC grant 'CORPHO'. C.C. is member of Institut Universitaire de France.

%
\section{APPENDIX}
In this appendix we use the exact solution \cite{DW} of the single-cavity problem to prove Eq.(\ref{rho_n}) and Eq.(\ref{b_lim_n}), holding for multiphotonic resonances in the limit $\gamma \ll F\ll \Delta\omega$.

For an isolated cavity, the density matrix of the stationary state is known analytically \cite{DW}:
\begin{eqnarray}
\rho_{n,m} &=& \frac{1}{\sqrt{n!m!}}\left(\frac{-2F}{U}\right)^n\left(\frac{-2F^{*}}{U}\right)^m 
\frac{\Gamma(c)\Gamma(c^{*})}{\Gamma(c+n)\Gamma(c^{*}+m)}\nonumber\\
&&\frac{\mathcal{F}(c+n,c^{*}+m,4|F/U|^2)}{\mathcal{F}(c,c^{*},8|F/U|^2)} \label{rhoan},
\end{eqnarray}
with
\begin{equation}\label{defc}
c = \frac{2(-\Delta \omega -i\gamma/2)}{U}. 
\end{equation}
In Eq.(\ref{rhoan}) $\Gamma(z)$ is the gamma special function which has poles at negative and zero integer values,
whereas $\mathcal{F}$ is an hypergeometric series given by:
\begin{equation}\label{hyp}
\mathcal{F}(c,d,z) = \sum_{k=0}^{\infty} \frac{\Gamma (c) \Gamma (d)}{\Gamma(c+k) \Gamma(d+k)}\frac{z^{k}}{k!}.
\end{equation}

In correspondence of a $q$-photon resonance, with $q>1$, the constant $c$ in Eq.(\ref{defc}) 
is given by
\begin{equation}
c=-(q-1)(1+i \frac{\gamma}{2 \Delta\omega}),
\end{equation}
implying that $c\approx -(q-1)$ for $\gamma \ll \Delta\omega$. As a result the quantities $\Gamma(c+k),\Gamma(c^*+k)$ are diverging for $0\leq k <q$, implying that certain coefficients of the hypergeometric series (\ref{hyp}) will actually diverge in the limit $\gamma \ll F \ll \Delta \omega$.

From the above consideration, the leading contribution in the two hypergeometric functions in Eq.(\ref{rhoan}) are
given by
\begin{equation}\label{eqF1}
\mathcal{F}(c,c^*,2z)\simeq \frac{\Gamma (c) \Gamma (c^*)}{\Gamma(c+q) \Gamma(c^*+q)} \frac{(2z)^q}{q!}
\end{equation} 
and 
\begin{equation}\label{eqF2}
\mathcal{F}(c+k,c^*+k,z)\simeq \frac{\Gamma (c+k) \Gamma (c^*+k)}{\Gamma(c+q) \Gamma(c^*+q)} \frac{z^{q-k}}{(q-k)!},
\end{equation}
respectively, where $z=4F^2/U^2$.

Substituting Eqs.(\ref{eqF1}) and (\ref{eqF2}) into the general expression for the density matrix, Eq.(\ref{rhoan}), we find
\begin{equation}
\rho_{kk}=\frac{1}{2^q k!}\frac{q!}{(q-k)!},
\end{equation}
which corresponds to Eq.(\ref{rho_n}) by replacing $q$ with $n$.

It is also easy to see that off-diagonal terms will instead vanish in the same limit $\gamma \ll F \ll \Delta \omega$.
Indeed, a similar analysis gives for $m<n$ (the opposite case can be treated in the same way):
\begin{equation}
\rho_{nm}=\frac{1}{\sqrt{n!m!}}\left(\frac{-2F}{U}\right)^{n-m} \frac{1}{2^q} \frac{q!}{(q-m)!  (n-m)!},   
\end{equation}
which indeed vanishes for vanishing pump amplitude, $F\rightarrow 0$.

The expression for the bosonic coherence is obtained in a similar way. The general formula for $\langle b \rangle$ is:
\begin{equation}
\langle b \rangle = \frac{F}{\Delta \omega+i\gamma/2}\times\frac{\mathcal{F}(1+c,c^{*},8|\frac{F}{U}|^2)}{\mathcal{F}(c,c^{*},8|\frac{F}{U}|^2)},
\end{equation}
 and the leading term in the geometric function appearing in the numerator is:
 \begin{equation}
\mathcal{F}(c+1,c^*,2z)\simeq \frac{\Gamma (c+1) \Gamma (c^*)}{\Gamma(c+1+q) \Gamma(c^*+q)} \frac{(2z)^q}{q!}.
 \end{equation} 
Using Eq.(\ref{eqF1}), we find:
\begin{equation}
\langle b \rangle = -(q-1)\frac{F}{\Delta \omega}.
\end{equation}


\begin{thebibliography}{0}
\expandafter\ifx\csname natexlab\endcsname\relax\def\natexlab#1{#1}\fi
\expandafter\ifx\csname bibnamefont\endcsname\relax
  \def\bibnamefont#1{#1}\fi
\expandafter\ifx\csname bibfnamefont\endcsname\relax
  \def\bibfnamefont#1{#1}\fi
\expandafter\ifx\csname citenamefont\endcsname\relax
  \def\citenamefont#1{#1}\fi
\expandafter\ifx\csname url\endcsname\relax
  \def\url#1{\texttt{#1}}\fi
\expandafter\ifx\csname urlprefix\endcsname\relax\def\urlprefix{URL }\fi
\providecommand{\bibinfo}[2]{#2}
\providecommand{\eprint}[2][]{\url{#2}}

\end{thebibliography}


\begin{thebibliography}{99}
\bibitem{Fisher}M. P. A. Fisher, P. B. Weichman, G. Grinstein, and D. S. Fisher, Phys. Rev. B {\bf40}, 546 (1989).
\bibitem{Greiner}M. Greiner, O. Mandel, T. Esslinger, T. W. H\"ansch and I. Bloch, Nature {\bf415}, 39 (2002)
\bibitem{Houck} A. A. Houck, H. E. T\"ureci, and J. Koch, Nature Phys. {\bf 8}, 292 (2012).
\bibitem{Deveaud} B. Deveaud (ed.),  {\it The Physics of Semiconductor Microcavities}, (Wiley-VCH, 2007).
\bibitem{RMP}I. Carusotto and C. Ciuti, Rev. Mod. Phys. {\bf 85}, 299 (2013).
\bibitem{Amo09}A. Amo {\it et al.}, Nature Phys. {\bf 5}, 805 (2009).
\bibitem{Tanese2012} D. Tanese, D. D. Solnyshkov, A. Amo, L. Ferrier, E. Bernet-Rollande, E. Wertz, I. Sagnes, A. Lema\^itre, P. Senellart, G. Malpuech, and J. Bloch.
Phys. Rev. Lett. {\bf 108}, 036405 (2012). 
\bibitem{Nardin} G. Nardin {\it et al.}, Nature Phys. {\bf 7}, 635 (2011).
\bibitem{Amo11} A. Amo {\it et al.}, Science  { \bf 332}, 6034 (2011).
\bibitem{Pigeon} S. Pigeon, I. Carusotto, and C. Ciuti, Phys. Rev. B {\bf 83}, 144513 (2011).
\bibitem{Sanvitto} D. Sanvitto {\it et al.}, Nature Phot. {\bf 5}, 610 (2011).
\bibitem{Imamoglu} A. Imamogl\v u, H. Schmidt, G.Woods, and M. Deutsch, Phys. Rev. Lett. {\bf 79}, 1467 (1997).
\bibitem{Fink} J.M. Fink, M. G\"oppl, M. Baur, R. Bianchetti, P. J. Leek, A. Blais, and A.Wallraff, Nature {\bf 454}, 315 (2008).
\bibitem{Liew} T.C. H. Liew and V. Savona, Phys. Rev. Lett. {\bf104},183601 (2010).
\bibitem{Bamba} M. Bamba, A. Imamoglu, I. Carusotto, and C. Ciuti, Phys. Rev. A  {\bf 83}, 021802 (R) (2011).
\bibitem{Hartmann06} M. J. Hartmann, F. G. S. L. Brandao and M. B. Plenio, Nature Physics {\bf 2}, 849 (2006).
\bibitem{Greentree} A. D. Greentree, C. Tahan, J. H. Cole and L. C. L. Hollenberg, Nature Physics {\bf 2}, 856 (2006).
\bibitem{Angelakis}D. G. Angelakis, M. F. Santos, and S Bose Phys. Rev. A { \bf 76}, 031805 (2007).
\bibitem{Wu} C.W. Wu,  M. Gao, Z.-J. Deng, H.-Y. Dai, P.-X. Chen, and C.-Z. Li, Phys. Rev. A {\bf 84}, 043827 (2011).
\bibitem{Tomadin} A. Tomadin, V. Giovannetti, R. Fazio, D. Gerace, I. Carusotto, H. E. T\"ureci, and A. Imamoglu Phys. Rev. A {\bf81}, 061801(R) (2010).
\bibitem{Carusotto09} I. Carusotto, D. Gerace, H. E. T\"ureci, S. De Liberato, C. Ciuti, and A. Imamo\v glu, Phys. Rev. Lett. {\bf 103}, 033601 (2009).
\bibitem{Ferretti} S. Ferretti, L. C. Andreani, H. E. T\"ureci, and Dario Gerace, Phys. Rev. A  {\bf 82}, 013841 (2010).
\bibitem{Hartmann10} M. J. Hartmann, Phys. Rev. Lett. {\bf 104}, 113601 (2010).
\bibitem{Leib10} M. Leib and M. Hartmann, New J. Phys. {\bf12}, 093031(2010).
\bibitem{Hafezi2011}  M. Hafezi, D. E. Chang, V. Gritsev, E. A.  Demler, M.D  Lukin,  Europhys. Lett.{\bf  94}, 54006 (2011).
\bibitem{Nissen}F. Nissen, S. Schmidt, M. Biondi, G. Blatter, H. E. T\"ureci, and J. Keeling, Phys. Rev. Lett. {\bf108}, 233603 (2012).
\bibitem{Carusotto2012} R. O. Umucalilar and I. Carusotto,
Phys. Rev. Lett. {\bf 108}, 206809 (2012).
\bibitem{Joshi} C. Joshi, F. Nissen, and J. Keeling, Phys. Rev. A {\bf 88}, 063835 (2013).
\bibitem{Jin} J. Jin, D. Rossini, R. Fazio, M. Leib, and M. J. Hartmann, Phys. Rev. Lett. {\bf 110}, 163605 (2013).
\bibitem{Angelakis2013} T. Grujic, S. R. Clark, D. Jaksch, and D. G. Angelakis
Phys. Rev. A {\bf 87}, 053846 (2013). 
\bibitem{PRL2013}A. Le Boit\'e, G. Orso, and C. Ciuti, Phys. Rev. Lett. {\bf 110}, 233601 (2013).

\bibitem{DW} P. D. Drummond and D. F. Walls, J. Phys. A {\bf13}, 725 (1980).
\bibitem{Vidanovic} I. Vidanovi\'c, Daniel Cocks, and Walter Hofstetter, Phys. Rev. A {\bf 89}, 053614 (2014).
\bibitem{Rydberg} A. Hu, T. E. Lee, and C. W. Clark, Phys. Rev. A {\bf 88}, 053627 (2013).

\end{thebibliography}
\end{document}